\documentclass[10pt,conference]{IEEEtran}
\usepackage{cite}
\usepackage{amsmath,amssymb,amsfonts}
\usepackage{algorithmic}
\usepackage{graphicx}
\usepackage{textcomp}
\usepackage{subfigure}
\usepackage{threeparttable}
\usepackage{float}
\usepackage{epsfig}
\usepackage{epstopdf}
\usepackage{amsmath}
\usepackage{multirow}
\usepackage{mathrsfs}
\usepackage{bm}
\usepackage[cmintegrals]{newtxmath}

\def\BibTeX{{\rm B\kern-.05em{\sc i\kern-.025em b}\kern-.08em
    T\kern-.1667em\lower.7ex\hbox{E}\kern-.125emX}}
\title{Coverage Analysis of Cellular-Connected\\ UAV Communications with 3GPP Antenna and Channel Models}
\author{\IEEEauthorblockN{Zhuangzhuang~Cui\IEEEauthorrefmark{1}\IEEEauthorrefmark{3}\IEEEauthorrefmark{4}, Ke~Guan\IEEEauthorrefmark{1}\IEEEauthorrefmark{4}, \.Ismail~G\"uven\c c\IEEEauthorrefmark{2}, Claude Oestges\IEEEauthorrefmark{3}, and Zhangdui~Zhong\IEEEauthorrefmark{1}\IEEEauthorrefmark{4}}
\IEEEauthorblockA{\IEEEauthorrefmark{1}State Key Lab of Rail Traffic Control and Safety, Beijing Jiaotong University, Beijing, China\\
\IEEEauthorrefmark{2}Department of Electrical and Computer Engineering, North Carolina State University, NC, USA\\
\IEEEauthorrefmark{3}ICTEAM, Universit\'e catholique de Louvain, Louvain-la-Neuve, Belgium\\
\IEEEauthorrefmark{4} Beijing Engineering Research Center of High-Speed Railway Broadband Mobile Communications, Beijing, China\\
Corresponding author: Ke Guan, Email: \texttt{kguan@bjtu.edu.cn}}}

\begin{document}

\maketitle

\begin{abstract}
For reliable and efficient communications of aerial platforms, such as unmanned aerial vehicles (UAVs), the cellular network is envisioned to provide connectivity for the aerial and ground user equipment (GUE) simultaneously, which brings challenges to the existing pattern of the base station (BS) tailored for ground-level services. Thus, we focus on the coverage probability analysis to investigate the coexistence of aerial and terrestrial users, by employing realistic antenna and channel models reported in the 3rd Generation Partnership Project (3GPP). The homogeneous Poisson point process (PPP) is used to describe the BS distribution, and the BS antenna is adjustable in the down-tilted angle and the number of the antenna array. Meantime, omnidirectional antennas are used for cellular users. We first derive the approximation of coverage probability and then conduct numerous simulations to evaluate the impacts of antenna numbers, down-tilted angles, carrier frequencies, and user heights. One of the essential findings indicates that the coverage probabilities of high-altitude users become less sensitive to the down-tilted angle. Moreover, we found that the aerial user equipment (AUE) in a certain range of heights can achieve the same or better coverage probability than that of GUE, which provides an insight into the effective deployment of cellular-connected aerial communications.
\end{abstract}


\section{Introduction}
Wireless networks are facing an unprecedented transformation where heterogeneous networks are designed to support seamless connectivity for different types of user equipment, in which unmanned aerial vehicles (UAVs), as critical aerial platforms, need reliable and high-speed communications to perform designated tasks, such as air surveillance and video stream transmission \cite{a1}. However, current UAV communications mainly rely on point-to-point (P2P) transmission over an unlicensed band (e.g.,~2.4~GHz), which is of low data rate, unreliable, vulnerable to interference, difficult to legitimately monitor and manage, and can merely operate over very limited range \cite{b2}. As a promising solution to remedying these defects, cellular base stations (BSs) are expected to serve not only ground user equipment (GUE) but also aerial user equipment (AUE) \cite{b1}. Accordingly, cellular-connected UAV communications have drawn significant attention in the literature \cite{b3,b4,b5}.

In this context, the coexistence of aerial and terrestrial users in cellular networks becomes an essential issue. Unfortunately, the existing cellular infrastructure is dedicated to serving ground users. Generally, the BS antennas are deployed with sector antennas with a down-tilted setting, which is not conducive for aerial platforms that fly in high altitudes. Nonetheless, thanks to the considerable line-of-sight (LOS) probability for high-altitude users, the coverage may be maintained effectively, even though the BS antenna provides limited directional gain. Consequently, this issue deserves an in-depth analysis in terms of corresponding performance evaluations, which will crucially facilitate the functional design and deployment of ground BSs as well as aerial platforms.

In prior works, the coverage probability of cellular UAVs was comprehensively analyzed \cite{b6,b7,b8}. These studies typically conducted complex theoretical derivations with the main focus on modeling the BS distribution such as homogeneous Poisson point process (PPP) and simplifying the interference power by Laplace transform. However, the channel models used in the analysis are merely assumed as simple distance-dependent attenuation models ($d^{-\alpha}$) where $d$ is the link distance and $\alpha$ is the path loss exponent \cite{b6}. Such a model may highly deviate from actual channel conditions that are confirmed to be greatly related to the environment, the frequency, and the user height, etc \cite{b17, a2}. Besides, a few works utilize a directional sector antenna for ground BS. For the tractable derivation, the majority of existing works employ an antenna with constant gains for the main lobe and side lobe in a certain range of angles of departure (AoDs) \cite{a3}, and some works exclude the antenna model for the easy-to-handle analysis \cite{a4}.

For practical considerations, we concentrate on the more realistic antenna and channel models proposed by the 3rd Generation Partnership Project (3GPP). In 3GPP TR~36.777~\cite{b9}, the ground BS is suggested to support aerial vehicles that fly within an altitude of 300~m. For the BS antenna, the model in 3GPP TR~36.873 \cite{b10} is used in this study. We also refer to the latest channel model including the path loss model and the LOS probability model of 3GPP TR 38.901 \cite{a5} in our analysis. The main contributions are summarized as follows. \textit{i}) We use the practical channel and antenna models in the coverage probability analysis and investigate the impacts of various factors such as the antenna numbers, frequencies, and user heights, which can be directly applied in the engineering design. \textit{ii}) We derive the Laplace transform of interference power, which enables the coverage probability evaluation to proceed in a more tractable way. Moreover, we define the critical height of AUE, in which the performance of aerial users is equivalent to the ground users, which provides a reference to the height control for aerial platforms. \textit{iii}) One of the important findings has shown that the coverage probability of the high-altitude user is insensitive to the down-tilted angles of ground BS, which suggests that we can pay less effort to the design and optimization of down-tilted angles.

The rest of the paper is organized as follows. Section~II introduces the system model including network, antenna, and channel models, as well as the BS association strategy. Section~III provides the theoretical derivation and corresponding approximation for the coverage probability. In section~IV, we conduct various simulations aiming to provide deployment insights for improving the coverage performance of AUE. Finally, the conclusion is drawn in Section~V.

\begin{figure}[tbp]
  \centering
   {\includegraphics[width=3in]{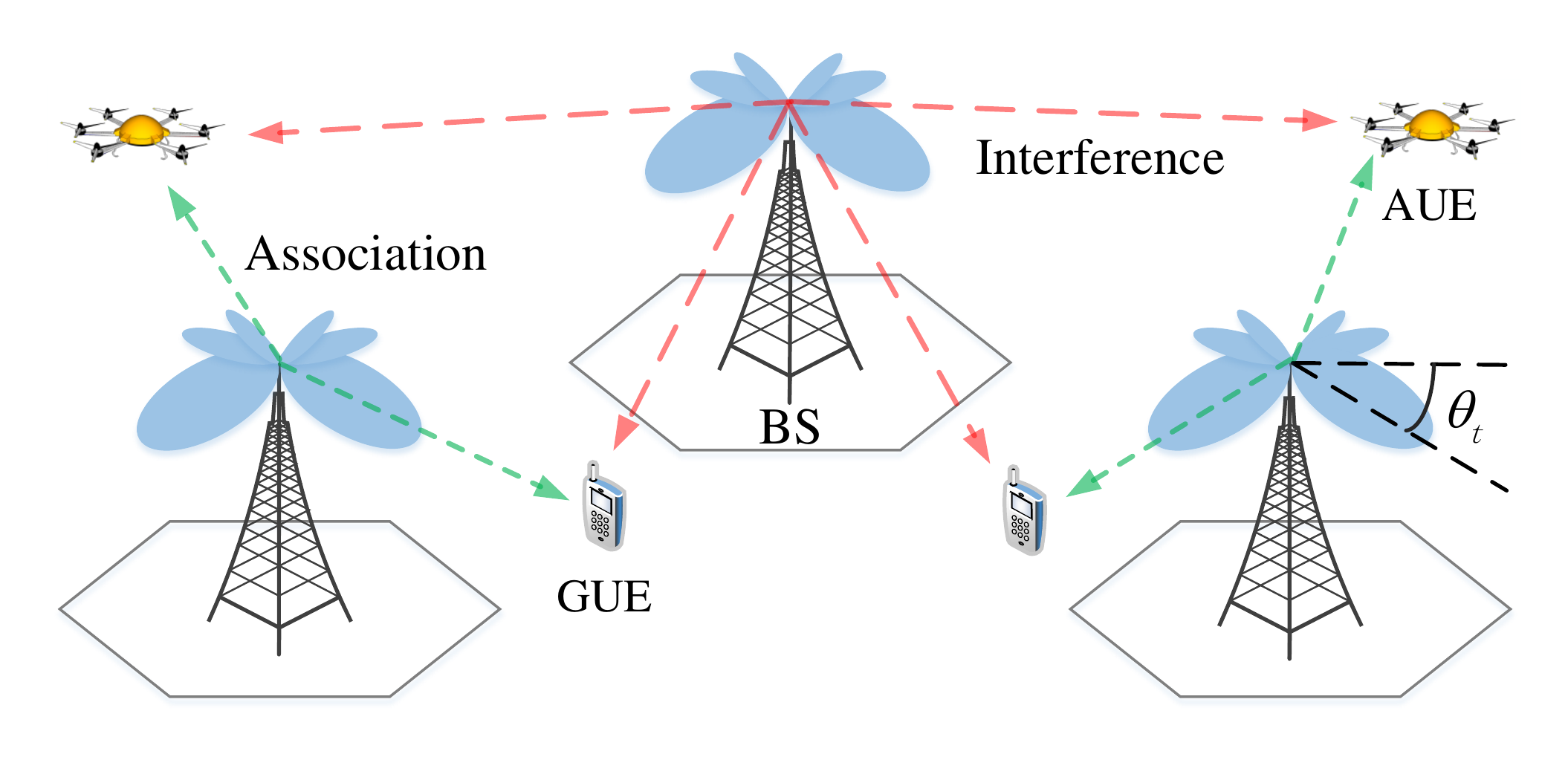}}
  \caption{Illustration of the AUE and GUE coexistence in cellular networks.}
   \label{Spawc}
\end{figure}
\section{System Model}

In this section, we will first present the system models composed of the antenna model, BS deployment model, and radio channel model. Then, the user association strategy and signal-to-interference-plus-noise ratio (SINR) will be introduced based on the presented system models.

\subsection{Network Model}
In this paper, we aim to study the downlink coverage probability of cellular-connected AUEs. The ground BSs are randomly distributed according to a homogeneous PPP $\Phi =\{b_i \in \mathbb{R}^2, \forall i \in \mathbb{N}\}$ with the fixed density $\lambda_{\text B}$ BSs/$\text{km}^2$ \cite{b11}. The BS antenna is adjustable in the element number of an array and down-tilted angle ($\theta_t$), meantime, the user antennas are omnidirectional. As shown in Fig.~1, the BS configured by directional sector antennas can serve AUE and GUE simultaneously. A typical user associates a single BS, and the other BSs will cause interference to the desired signal. The two-dimensional (2D) distance between the user and the $i$-th BS is denoted as $r_i$, and thus the three-dimensional (3D) distance is $d_{\text{3D}}=\sqrt{r_i^2+(h_{\text{BS}}-h)^2}$ where $h$ is the user height. In this paper, the height of GUE is assumed as 1.5~m, and the height of AUE ranges from 1.5~m to 300~m.

\subsection{BS Antenna Model}
The BS antenna is based on the technical report of 3GPP 36.873 \cite{b10}. The horizontal and vertical radiation pattern in dB can be respectively expressed as
 \begin{equation}
\bm{A}_{\text{E,H}}(\varphi)=-\min\left[12\left(\frac{\varphi}{\varphi_{3\text{dB}}}\right)^2,A_{\text m}\right],
 \end{equation}
  \begin{equation}
\bm{A}_{\text{E,V}}(\theta)=-\min\left[12\left(\frac{\theta-90}{\theta_{3\text{dB}}}\right)^2,SLA_{\text v}\right],
 \end{equation}
where $\theta$ is the elevation angle, defined between $0^{\circ}$ and $180^{\circ}$ ($90^{\circ}$ represents perpendicular to the array antenna aperture), and
$\varphi$ is the azimuth angle, defined between $-180^{\circ}$ and $180^{\circ}$. The $\varphi_{3\text{dB}}$ and $\theta_{3\text{dB}}$ are the horizontal and the vertical 3~dB beam-width, respectively. $A_{\text m}$ is the front-to-back ratio and $SLA_{\text v}$ is the side-lobe level limit.
The pattern of the element in the array can be represented as
  \begin{equation}
\bm{A}_{\text E}(\varphi, \theta)=G_{\text E,\max}-\min\{-[A_{\text{E,H}}(\varphi)+A_{\text{E,V}}(\theta)],SLA_\text{v}\},
 \end{equation}
where the $G_{\text E,\max}$ is the maximum element gain.

In addition, the array pattern can be obtained by
  \begin{equation}
  \begin{aligned}
\bm{A}_{\text A}(\varphi, \theta, \theta_t)= A_{\text E}(\varphi, \theta) +AF(\varphi, \theta,\theta_t),
 \end{aligned}
  \end{equation}
  \begin{equation}
  \begin{aligned}
AF(\varphi, \theta,\theta_t)=10\log_{10}[1+\rho(|\bm{v}\cdot \bm{w}^T|^2-1)],
  \end{aligned}
  \end{equation}
    \begin{equation}
\bm{v}=[v_{1,1},v_{1,2}...,v_{1,N_{\text V}};...;v_{N_{\text H},1},v_{N_{\text H},2}...,v_{N_{\text H},N_{\text V}}]^T,
 \end{equation}
  \begin{equation}
\bm{w}=[w_{1,1},w_{1,2}...,w_{1,N_{\text V}};...;w_{N_{\text H},1},w_{N_{\text H},2}...,w_{N_{\text H},N_{\text V}}]^T,
 \end{equation}

where  $\theta_{t}$ is the electronic tilt angle and $\rho$ is the correlation level between elements. Besides, the array factor of the planar array can be represented by $\bm{\widetilde{w}}=\bm{w}\cdot \bm{v}$ with $v_{m,n}=\exp{(i\cdot 2\pi(n-1)\psi_v+(m-1)\psi_h)}$, $\psi_v=\frac{d_{\text V}}{\lambda}\cos(\theta)$, and $\psi_h=\frac{d_{\text H}}{\lambda}\sin(\theta)\sin(\varphi)$ where $\lambda$ is the wavelength. Note that $m=1,2,...,N_{\text H} and n=1,2,...,N_{\text V}$. Herein $\bm{w}$ is the weighting factor, which can provide control of side lobe levels and can also provide both horizontal and vertical electrical steering, which is given by $w_{m,n}=\frac{1}{\sqrt{N_{\text H} N_{\text V}}}\exp((i \cdot 2\pi((n-1)\psi'_v+(m-1)\psi'_h)$ and $\psi'_v=\frac{d_{\text V}}{\lambda}\sin(\theta_{t})$ and $\psi'_h=\frac{d_{\text H}}{\lambda}\cos(\theta_{t})\sin(\varphi_{s})$. Note that $d_{\text V}$ and $d_{\text H}$ is the element space in the vertical and horizontal dimensions, respectively, which are generally considered as 0.5$\lambda$ or 0.8$\lambda$. In this paper, we exploit a uniformly distributed linear array (ULA) with $N_{\text V}=N$, $d_{\text V}=0.5\lambda$ and $N_{\text H}=1$.

\subsection{Channel Model}
The channel models for cellular-connected aerial vehicles were originally proposed in the 3GPP TR 36.777 \cite{b9}. With the development of 3GPP TR 36.873 \cite{b10} and 3GPP TR 38.901 \cite{a5}, these models are gradually complete. The models are suggested to be used for a maximum altitude of 300~m because of many practical considerations such as flight safety and legislation. Hence, we integrate the models proposed in above 3GPP reports and employ them in the assessment. Given the long-distance flight of AUE, we consider an urban macro (UMa) scenario, where the path loss in the LOS condition for the user height in 1.5-22.5~m can be calculated by
   \begin{equation}
PL_{\text{L}}= \begin{cases}
28.0+22\log_{10}(d_{\text{3D}})+20\log_{10}(f_c); r_i\in [10~\text{m}, d_{\text B})\\
28.0+40\log_{10}(d_{\text{3D}})+20\log_{10}(f_c)\\-9\log_{10}((d_{\text B})^2+(h_{\text{BS}}-h)^2); r_i\in [d_{\text B}, 5~\text{km}]
\end{cases}
   \end{equation}
where $d_{\text{3D}}$ is in~m and $f_c$ is in GHz. The break distance $d_{\text{B}}$ is equal to $4h_{\text{BS}}hf_c/c$, where $c$ is the speed of light. For the non-LOS (NLOS) case, the path loss can be calculated by
\begin{equation}
PL_{\text{N}}= \max{\{PL_{\text{L}}, PL'_{\text{N}}\}},
\end{equation}
where $PL'_{\text{N}}$ can be calculated by
\begin{equation}
  \begin{aligned}
PL'_{\text{N}}&=13.54+39.08\log_{10}(d_{\text{3D}})+20\log_{10}(f_c)\\&-0.6(h-1.5).
\end{aligned}
   \end{equation}

For $h\in (22.5~\text{m}, 300~\text{m}]$ and $d_{\text{2D}}\leq 4~\text{km}$, the path loss in the LOS condition is given by \begin{equation}
PL_{\text{L}}= 28.0+22\log_{10}(d_{\text{3D}})+20\log_{10}(f_c).
\end{equation}

For $h \in (22.5~\text{m}, 100~\text{m}]$ and $d_{\text{2D}}\leq 4~\text{km}$, the path loss for NLOS can be calculated by
   \begin{equation}
     \begin{aligned}
PL_{\text{N}}&= -17.5+(46-7\log_{10}(h))\log_{10}(d_{\text{3D}})\\&+20\log_{10}(40\pi f_c/3).
\end{aligned}
  \end{equation}

The LOS probability also plays a vital role in determining the total loss \cite{b13}. In \cite{b9}, the LOS probability is as the function of the user height and the 2D distance. For $h\in [1.5~\text{m}, 22.5~\text{m}]$, the probability is given by
   \begin{equation}
\mathbb{P}_{\text{L}}= \begin{cases}
1, & r_i\leq 18~\text{m}\\
\left[\frac{18}{r_i}+\exp\left(-\frac{r_i}{36}\right)\left(1-\frac{18}{r_i}\right)\right]\times\\ \left(1+C'h\frac{5}{4}\left(\frac{r_i}{100}\right)^3\exp\left(-\frac{r_i}{150}\right)  \right), &  r_i\geq 18~\text{m}
\end{cases}
   \end{equation}
where $C'h=0$ when $h \leq  13~\text{m}$ and is equal to $\left((h-13)/10\right)^{1.5}$ for $h \in (13~\text{m}, 22.5~\text{m}]$.

For the short distance ($\leq d_1$) and the altitude larger than 100~m, the LOS probability is equal to 1 according to the 3GPP model. For $r_i\geq d_1$ and $h\in (22.5~\text{m}, 100~\text{m}]$, the LOS probability is expressed as
   \begin{equation}
\mathbb{P}_{\text{L}}= \frac{d_1}{r_i}+\exp\left(-\frac{r_i}{p_1}\right)\left(1-\frac{d_1}{r_i}\right),
   \end{equation}
where $p_1$ and $d_1$ are altitude-dependent parameters with $p_1=4300\log_{10}(h)-3800$ and $d_1=\max((460\log_{10}(h)-700),18)$.

Thus, the total propagation loss can be expressed as
\begin{equation}
 P(r_i,h)=PL_{\text L}(r_i,h)\mathbb{P}_{\text L}(r_i,h)+PL_{\text N}(r_i,h)(1-\mathbb{P}_{\text L}(r_i,h)).
 \end{equation}

Besides, the small-scale fading is necessary to be incorporated in the evaluation. For the generality, the Nakagami-$m$ fading model is used in this paper. Accordingly, the channel gain $|g_i|^2$ follows a gamma distribution with the probability density function (PDF) expressed as [17] with $f_{|g_i|^2}(x)=\frac{m^{m}x^{m-1}}{\Gamma(m)}\exp(-mx)$ where $\Gamma(m)$ is the gamma function given by $\Gamma(m)=\int_0^{\infty}x^{m-1}\exp(-x)dx$. We assume that the small-scale fading of associated link and interfering links is independent and identically distributed (i.i.d.) with the same $m$.

\subsection{BS Association and SINR}
For the user association scheme, the closest distance strategy is commonly used as\cite{b14}. We denote the coordinate of typical user as $(x_{\text u},y_{\text u},h)$ in Cartesian coordinate system, and thus the 2D horizontal distance between BS and typical UE is $r_i=\sqrt{(x_i-x_{\text u})^2+(y_i-y_{\text u})^2}$ where $i$ represents the $i$-th BS with the coordinate $(x_i,y_i,h_{\text{BS}})$. %Then, by the PPP assumption, we have the PDF of $r_0$ as $f_{r_0}(x)=2\pi\lambda_\text{B} x \exp(-\pi\lambda_\text{B} x^2)$.

Afterwards, the instantaneous SINR of the typical user associated with a BS located at point $b_0$ is given by
\begin{equation}
\text{SINR} =\frac{P_t P(r_0,h) G(\theta_0,\phi_0,\theta_t) |g_0|^2}{\sum_{i \in \Phi/\{b_0\}}P_t P(r_i,h) G(\theta_i,\phi_i,\theta_t) |g_i|^2+\sigma_n^2}, %I+\sigma_n^2
\end{equation}
where $P_t$ is the transmit power. $G(\theta_0,\phi_0,\theta_t)$ is the antenna gain with the down-tilted angle $\theta_t$. In particular, the aggregate interference $I$ can be written as
\begin{equation}
I=\sum_{i=1}^{N_{\text B}-1} P_t P(r_i,h) G(\theta_i,\phi_i, \theta_t )|g_i|^2,\; i \in \Phi/\{b_0\}
\end{equation}
where $N_{\text B}$ is the number of BS in a considered area. The elevation and azimuth angles can be determined by the geometric calculations that $\theta_i=\arctan\left(\frac{h_{\text{BS}}-h}{r_i}\right)$ and $\phi_i=\arctan\left(\frac{x_i-x_{\text u}}{y_i-y_{\text u}}\right)$ where we assume $x_{\text u}=y_{\text u}=0$.

\section{Coverage Analysis}
In this section, we will first give the method of evaluating the coverage probability based on approximating the interference power. Then, we will define a critical height for AUE, which provides a useful reference for the height operation.

\subsection{Coverage Probability Evaluation}
 The coverage probability is defined as the probability that the received SINR exceeds a given threshold $T$ \cite{b14}. The calculation can be conducted by
\begin{equation}
     \begin{aligned}
  P_{\text{cov}}(T)&=\mathbb{P}[\text{SINR}  > T]=\mathbb{E}_{\Phi}[\mathbb{P}(\text{SINR} (x)>T)],
    \end{aligned}
\end{equation}
where $x$ is a considered variable such as the height ($h$), the down-tilted angle ($\theta_t$), or the number of antenna element ($N$).

Since the small-scale fading follows Nakagami-$m$ distribution, we can first determine $\mathbb{P}(\text{SINR}>T)$ as follows,
 \begin{equation}
     \begin{aligned}
 &\mathbb{P}(\text{SINR}>T)= \mathbb{P}\left[|g_0|^2>s(I+\sigma_n^2)\right],
 \\&\overset{(a)}{=} \mathbb{E}_I \left[\gamma\left(m, ms(I+\sigma_n^2)\right)/\Gamma(m)\right],
 \\&\overset{(b)}{=} \mathbb{E}_I \left[\sum_{k=0}^{m-1} \frac{\left(ms(I+\sigma_n^2)\right)^k}{k!}\exp(-ms\sigma_n^2)\exp(-msI)\right],\\ &\overset{}{=}\sum_{k=1}^{m}(-1)^{k+1}\binom{m}{k}\exp(-ms\sigma_n^2)\exp(-msI),
   \end{aligned}
\end{equation}
where $s= T(P_t P(r_0,h) G(\theta_0,\phi_0,\theta_t))^{-1}$. Moreover, ($a$) follows the complementary cumulative distribution function (CCDF) of gamma random variable $|g_0|^2$, ($b$) leverages the definition of incomplete gamma function with parameter $m$. $\mathcal{L}_I(ms)$ is the Laplace transform of interference power $I$, which can be given by \cite{a3}
\begin{equation}
   \begin{aligned}
  &\mathcal{L}_I(ms)=\mathbb{E}_I[\exp(-msI)]\\&=\mathbb{E}_{\Phi,|g_i|^2}\left[\exp\left(-ms\sum_{i \in\Phi/\{b_0\}} p(r_i)|g_i|^2\right)\right],
 \\&\overset{}{=}\mathbb{E}_{\Phi}\left[\prod_{i \in \Phi/\{b_0\}} \mathbb{E}_{|g_i|^2}[\exp(-msp(r_i) |g_i|^2)]\right],
 \\&\overset{(d)}{=}\exp\left(-2\pi \lambda_\text{B} \int_{r_0}^\infty \left(1-\left(1+\frac{p(r_i)T}{mp(r_0)}\right)^{-m}\right) r_idr_i\right),
   \end{aligned}
 \end{equation}
 where $p(r_i)=P_t P(r_i,h) G(\theta_i,\phi_i,\theta_t)$ and ($d$) follows the probability generating funationn (PGFL) of PPP $\mathbb{E}[\prod_{x \in \phi} f(x) ] = \exp(-\lambda \int_{\mathbb{R}^2} (1-f(x))dx)$ \cite{b18}. Notably, an exact expression of coverage probability is difficult to obtain because the path loss and antenna models are intermittent functions of $r_i$ and hardly tractable in an analytical way, we can make use of the summation instead of integral calculation to deal with the interference power. As a consequence, the approximate coverage probability is expressed as
 \begin{equation}
    \begin{aligned}
P_{\text{cov}} =&\sum_{k=1}^{m}(-1)^{k+1}\binom{m}{k}\exp(-ms\sigma_n^2)\times\\&\exp\left(-2\pi\lambda_\text{B}\sum_{i=1}^{N_B-1}\left(1-\left(1+\frac{p(r_i)T}{mp(r_0)}\right)^{-m}\right)\right).
    \end{aligned}
 \end{equation}
 In particular, for $m=1$, the coverage probability is given by
 \begin{equation}
    \begin{aligned}
 P_{\text{cov}}& =\exp\left(\frac{-T\sigma_n^2}{p(r_0)}-2\pi \lambda_\text{B}\sum_{i=1}^{N_B-1}\left(\frac{p(r_i)T}{p(r_0)+p(r_i)T}\right) \right).
    \end{aligned}
 \end{equation}
\subsection{Critical Height of AUE}
For a feasible operation of AUE, it is pivotal to determine the critical height, at which the coverage probability of AUE can attain the same level as the GUE, which is crucial to the coexistence of AUE and GUE in cellular networks. Accordingly, we denote the critical height as $h_c$ that can be obtained by solving the following problem,
\begin{equation}
  P_{\text{cov}}(h_c)=  P_{\text{cov}}(1.5 \;\text{m}), \;\text{for given}  \; \{T, N, \theta_t, \lambda_{\text{B}}, f_c\},
\end{equation}
where $h_c$ can be obtained by numerical simulations. Note that the coverage probability first increases with the height of UE that ranges from 1.5~m to the height $(h_{\text{BS}}-r_0\tan(\theta_t))$ that the typical user aligns with the main lobe of BS antenna, hereafter, the probability will tend to decline with the height because of the decreasing antenna gain and the increasing path loss. Thus, the recommended height can be from 1.5~m to $h_c$.

  \begin{table}[tbp]
\centering
  \caption{Simulation Parameters for Coverage Evaluation } \label{table1}
  \begin{tabular}{|c|c|}
\hline
  \textbf{Parameter} &  \textbf{Value} \\
    \hline
Frequency ($f_c$) & 5, 28~GHz \\
  \hline
Transmit power ($P_t$)  &  25~dBm  \\
\hline
Noise power  ($\sigma_n^2$) &  -95~dBm  \\
\hline
BS height   ($h_{\text{BS}}$) & 25~m\\
\hline
PPP intensity  ($\lambda_{\text B}$) & 5 BSs/km$^2$ \\
\hline
UE height ($h$) & 1.5~m-300~m \\
\hline
Number of element  ($N$) & 16, 32, 64 \\
\hline
Down-tilt ($\theta_t$) & $5^\circ$, $10^\circ$, $15^\circ$  \\
\hline
\end{tabular}
\end{table}
\begin{figure}[tbp]
  \centering
  {\includegraphics[width=3in]{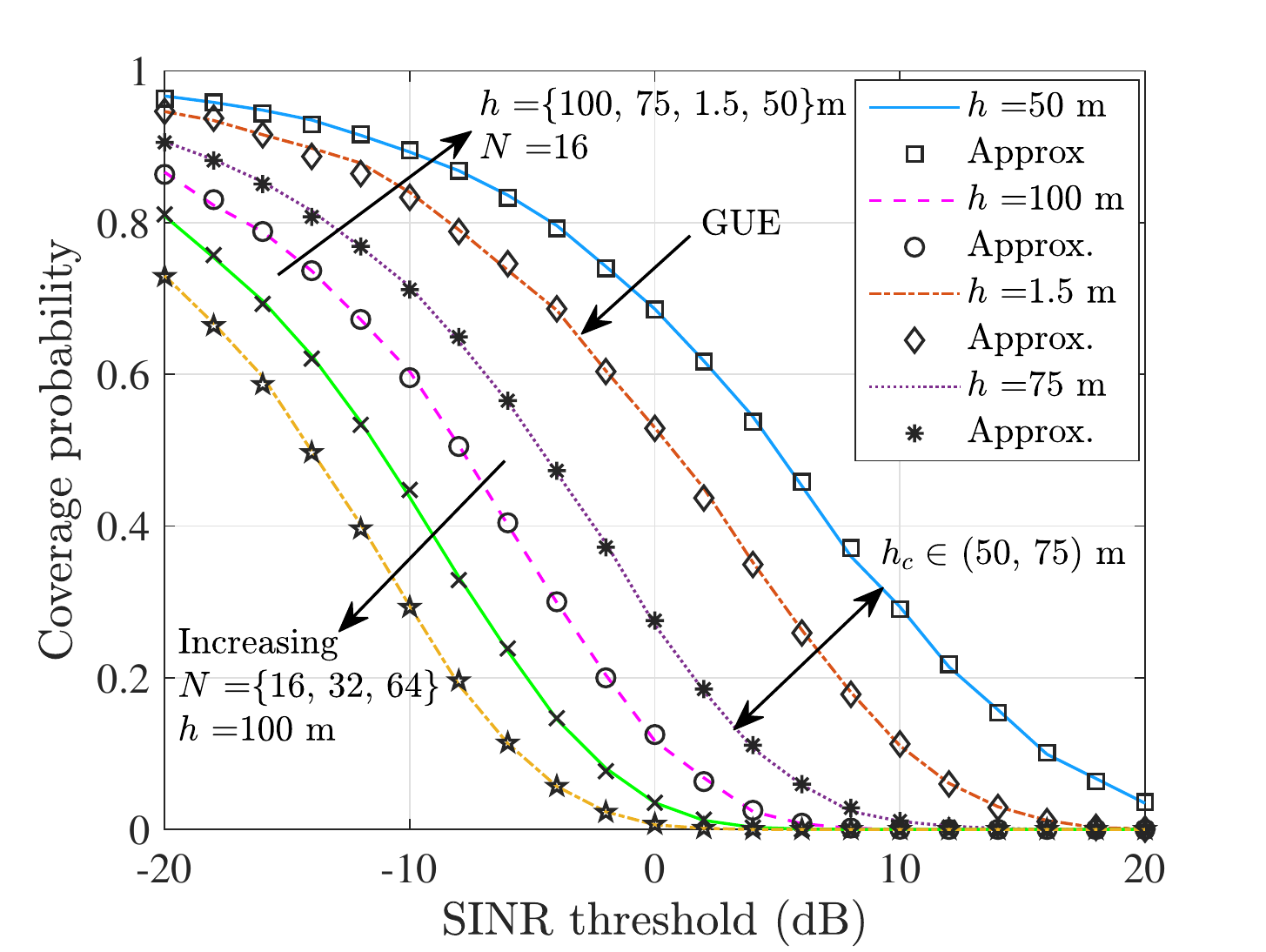}}
  \caption{Simulated (lines) and approximated (markers) coverage probability results for different user heights and element numbers of an antenna array.}
   \label{cp_validation}
\end{figure}
\section{Performance Evaluation}
In this section, we conduct numerous Monte-Carlo simulations with $10^4$ iterations, with parameters as summarized in Table~I. We first compare the approximations with simulated results to show the correctness of our derivation. Afterward, we focus on evaluating various factors that affect the coverage probability. The frequencies are set to 5~GHz and 28 GHz for potential sub-6~GHz and millimeter-wave applications \cite{b15}.
\subsection{Validation}
For the validation of our approximation, we plot the simulated (lines) and approximated (markers) results of coverage probability in Fig.~2. The results present ideal agreements between them for diverse user heights and antenna numbers, which shows the high accuracy of the derived approximation. Because results can always perfectly match with simulations, we only include the approximated result in the sequel.

\subsection{Impact of Element Numbers of ULA}
 In order to find the impact of the element number, $N$ is set to 16, 32, and 64 with fixed parameters ($\theta_t=5^\circ$, $f_c=28$~GHz) for the AUE at 100~m. As shown in Fig.~2, the larger number of elements gives rise to the smaller coverage probability, which can be well explained from the perspective of the beam-width of antenna. For the antenna model in 3GPP, the larger number of elements leads to the narrower beam-width, which results in more limited coverage by nature.
 \begin{figure}[tbp]
  \centering
  {\includegraphics[width=3in]{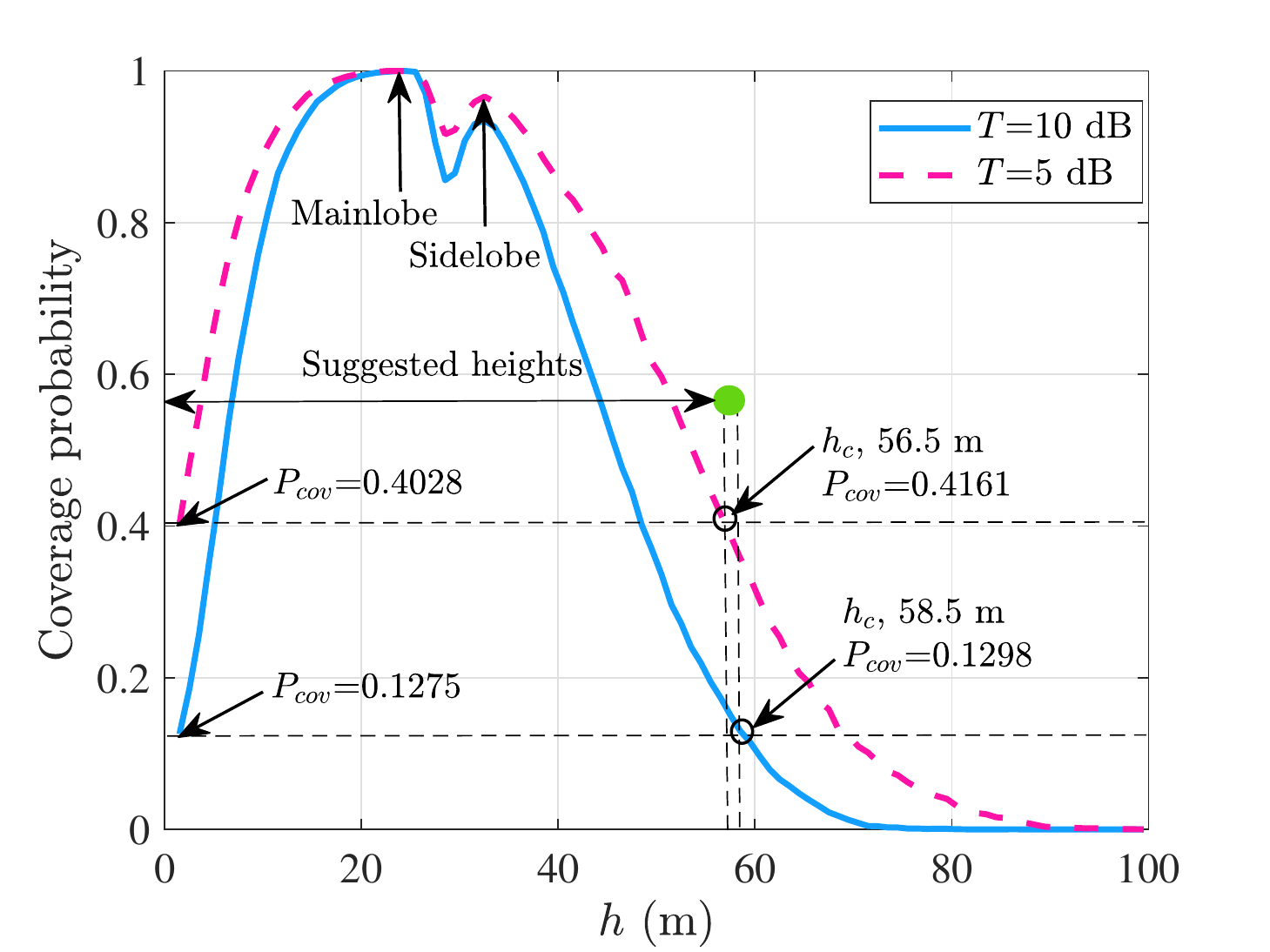}}
  \caption{Coverage probability versus user height for different SINR thresholds.}
   \label{CP_H_n}
\end{figure}
\begin{figure}[tbp]
  \centering
  \includegraphics[width=3in]{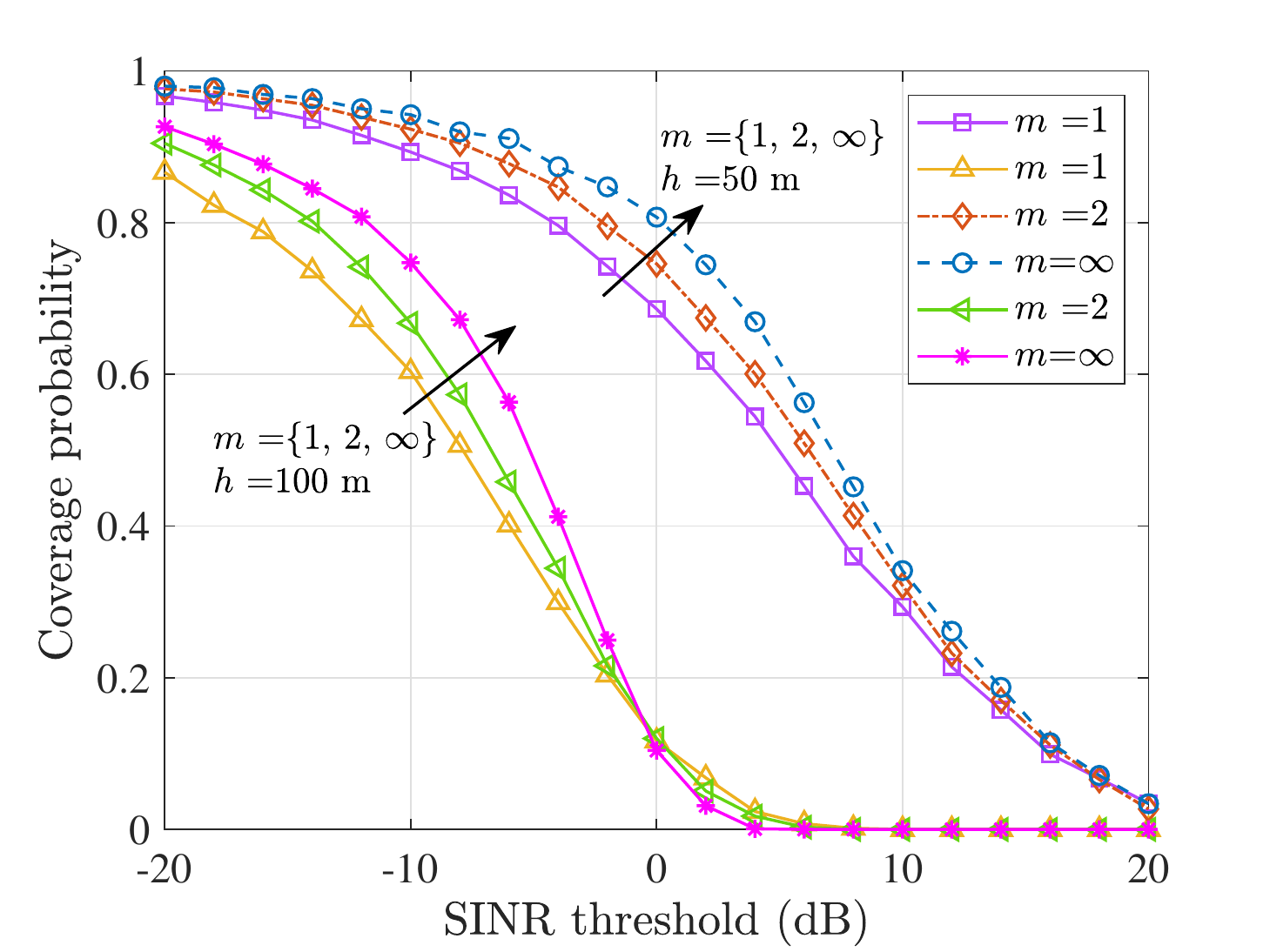}
  \caption{The impact of small-scale fading on coverage probability.}
   \label{ssf}
\end{figure}
\subsection{Impact of User Heights}
To evaluate the impact of user height on the coverage probability, we select four separated heights for GUE (1.5~m) and AUE (50~m, 75~m, 100~m). As shown in Fig.~2, the coverage probability for AUE at 100 m is the smallest among the four heights. Interestingly, even though the AUE is mainly served by the side lobe of the BS antenna, the coverage probability of AUE at 50~m is larger than that of GUE at 1.5~m, which is owing to the considerable LOS probability for the AUE in higher altitudes. It indicates that the SINR gain that the LOS probability brings has surpassed that the BS antenna produces at that height. Thus, for AUE higher than the BS, the LOS probability plays a more important role in the coverage probability than the impact of BS antenna, although the AUE is merely served by the low-gain side lobe and the GUE can obtain more gain from the main lobe. In addition, it is observed that the critical height ($h_c$) of AUE for current settings is somewhere between 50 and 75~m.

To further determine the critical height, we plot the coverage probability with respect to the user height, shown in Fig.~3. The results show that the critical heights are 56.5~m and 58.5~m for SINR thresholds of 5~dB and 10~dB, respectively. Accordingly, it is suggested that the AUE can be operated within the height of $h_c$ to acquire coverage performance that is not worse than that of GUE. In addition, to obtain better coverage for AUE, there are two options. As shown in Fig.~3, two peaks of coverage probability occur at specific heights, which correspond to the conditions that the main lobe and side lobe of antenna pattern straightly point to the AUE, respectively. In Fig.~3, the two heights are 24.5~m and 32.5~m. It is also shown that the different SINR thresholds will not change these heights because the heights are mainly determined by the antenna deployment. As aforementioned, the first height can be calculated as $h_{\text{BS}}-r_0\tan(\theta_t)$, that is approximately equal to $h_{\text{BS}}$ due to the small $\theta_t$. We have to mention that the coverage probability drops between two peaks, which is caused by the null pattern between the main and side lobe of BS antenna.

\begin{figure}[tbp]
  \centering
 \includegraphics[width=3in]{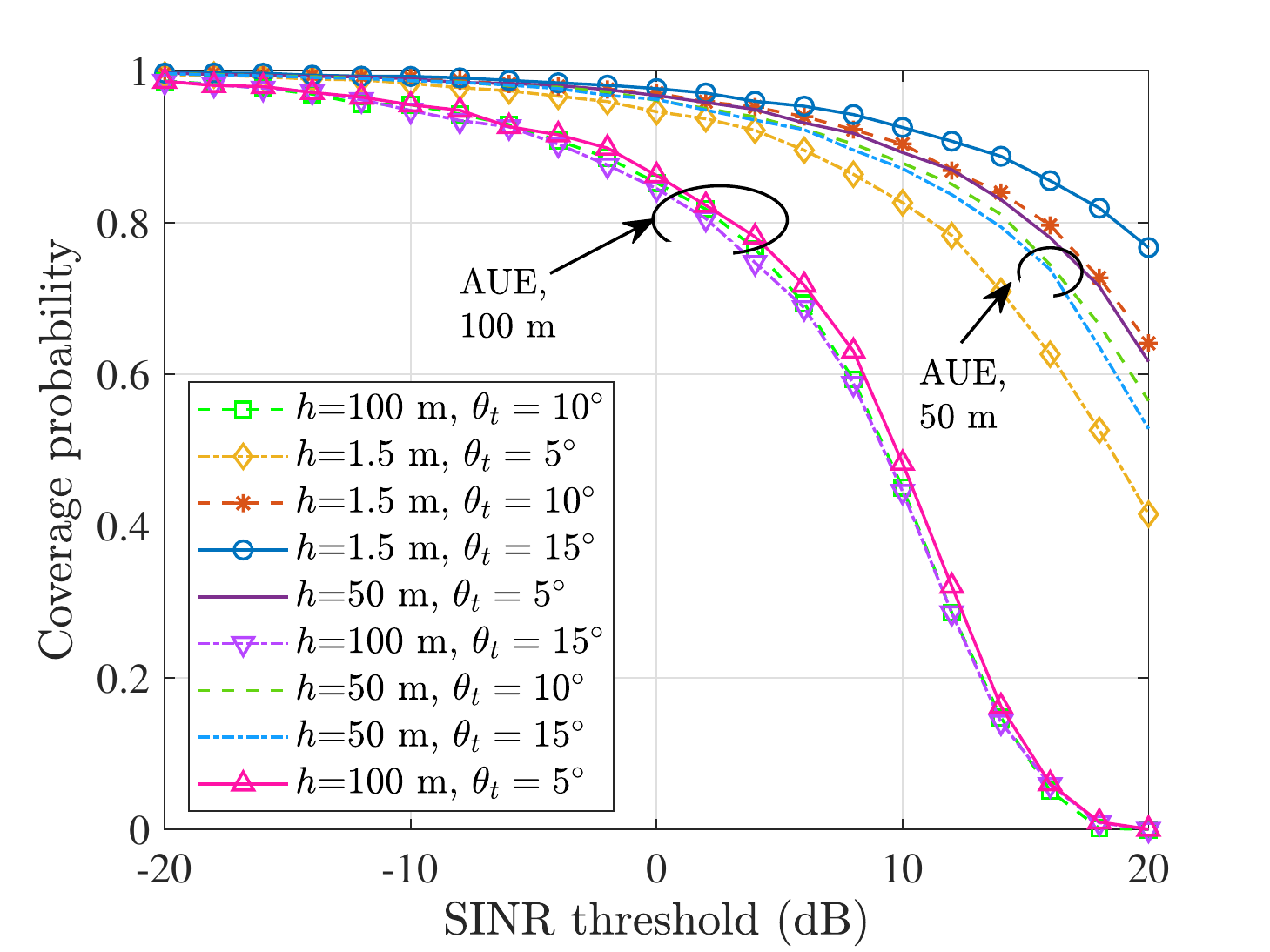}
  \caption{The impact of down-tilted angles on coverage probability.}
   \label{cp_theta}
\end{figure}
\begin{figure}[tbp]
  \centering
  \includegraphics[width=3in]{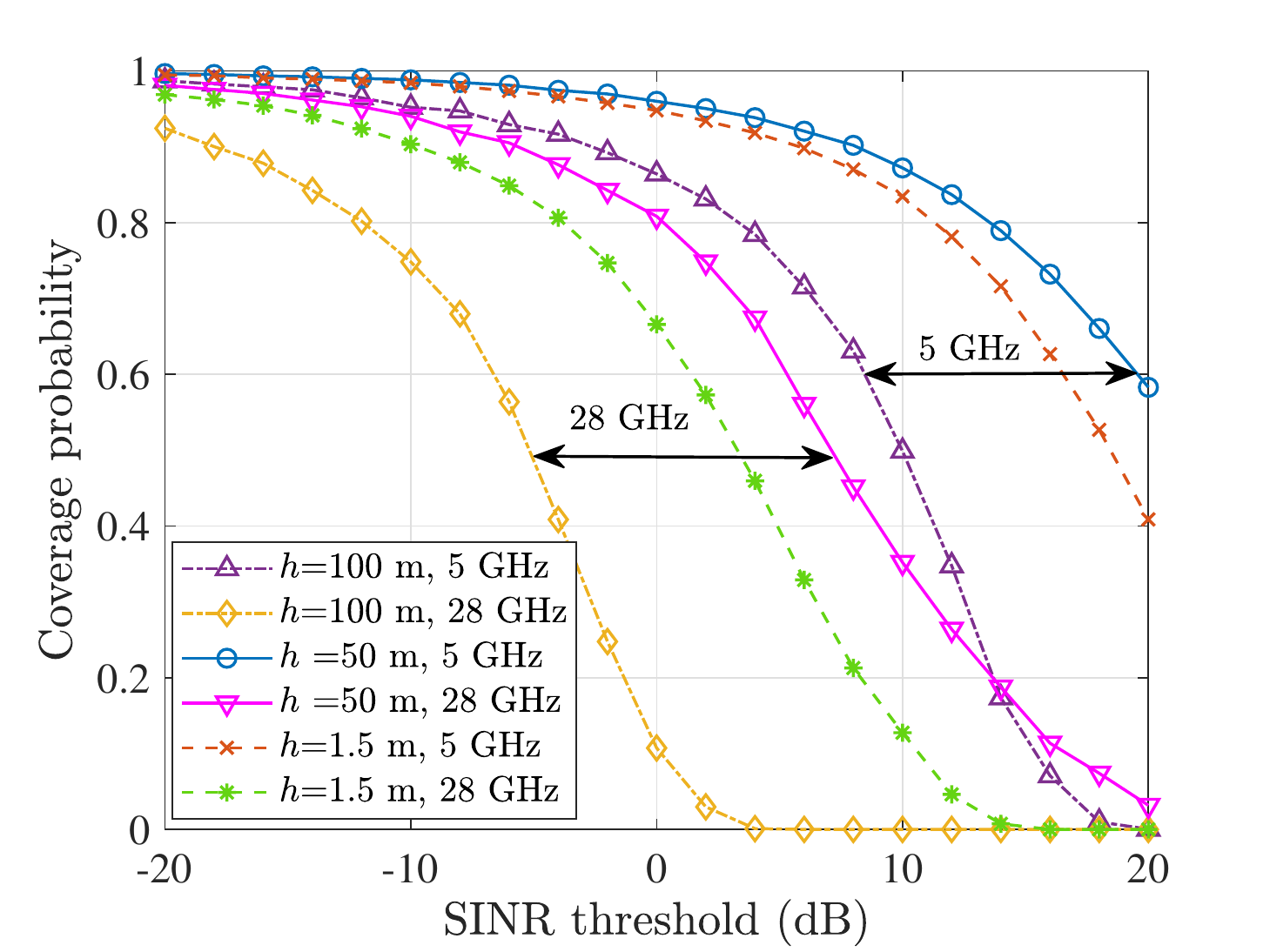}
  \caption{The impact of different frequencies on coverage probability.}
   \label{cp_h}
\end{figure}
\subsection{Impact of Small-Scale Fading}
We then look into the impact of small-scale fading on the coverage probability, shown in Fig.~4. It is known that the condition of $m=1$ is in accordance with the Rayleigh fading that represents the rich scattering. For $m\to\infty$, there is no fade existing. Thereby, the conditions of $m=1$ and $m\to\infty$ correspond to the lower and upper bounds of the impact of small-scale fading, respectively. As expected, the coverage probability increases with the increasing $m$. However, for large SINR thresholds ($\ge 10$~dB), the impact of $m$ becomes negligible since the instantaneous SINR is needed to exceed the large SINR threshold and thus mainly influenced by the large-scale fading, i.e., path loss.

\subsection{Impact of Down-tilted Angles}
It is known that the current BS is equipped with down-tilted antennas, which is naturally considered unfavorable for the AUE. However, we found the setting seems to have limited influences on the AUE. We compare the coverage probability for three down-tilted angles ($5^\circ, 10^\circ, 15^\circ$) for the GUE at 1.5~m and AUE at 50~m and 100~m. As shown in Fig.~5, the influence of down-tilted angles on the coverage probability of AUE is much less than that of GUE. For instance, for the SINR threshold of 10~dB, when $\theta_t$ increases from $5^\circ$ to $15^\circ$, the coverage probability increases 9.9\% for GUE and reduces 3.7\% and 2.1\% for AUE at 100~m and 50~m, respectively. The finding indicates that the coverage probability of high-altitude AUE is not sensitive to the down-tilted angles, which can be interpreted by the BS antenna and LOS probability. Since the side lobe has low gain, the SINR of AUE in high altitude is mainly dominated by the strong LOS condition, which leads to the insensitivity to down-tilted angles.

\subsection{Impact of Frequencies}
We finally compare the coverage probability for different frequencies. As shown in Fig.~6, it is obvious that the high frequency gives rise to the declining coverage probability due to the severe path loss. As an example, the coverage probability for AUE at 100~m can still achieve 50\% coverage for $T=10$~dB and $f_c=5$~GHz, however, it drops to 0 at 28~GHz. It is manifest that the impact of the frequency is considerable for the AUE. To robustly adapt to distinct frequencies, the AUE may need to adjust its height dynamically.

\section{Conclusion}
In this paper, we analyze the coverage probability for cellular-connected UAV communications by employing 3GPP antenna and channel models and comprehensively investigate an abundance of impacts of the user heights, antenna numbers, down-tilted angles, and frequencies, etc. Through the theoretical derivation of interference power, we develop a tractable way to evaluate the coverage probability. The perfect agreement between simulation and approximation confirms our correctness. In the evaluation, the impacts of various factors can be well explicated from the antenna, path loss, LOS probability, as well as small-scale fading. More invaluably, we provide many feasible solutions to the practical engineering design. First of all, the height control of the aerial platform can refer to our defined critical height, which can guarantee the performance of AUE is not worse than that of GUE. Then, we found that the down-tilted angles of the BS antenna have an insignificant effect on the coverage probability of AUE, which suggests that we should pay more effort to the design of array number and the selection of carrier frequency on the BS side. Overall, these findings on both AUE and BS sides obtained in the evaluation provide important references to their effective deployments in cellular-connected UAV communications.

\appendices

\section*{Acknowledgment}
This work was supported by the NSFC under Grant (61771036, 61911530260, U1834210, and 61725101), the NSF (CNS 1453678), the China Scholarship Council (CSC) (No. 202007090173), and the State Key Laboratory of Rail Traffic Control and Safety (Contract No. RCS2019ZZ005).

\ifCLASSOPTIONcaptionsoff
  \newpage
\fi


\begin{thebibliography}{99}
\bibitem{a1}
Y. Zeng, R. Zhang and T. J. Lim, ``Wireless communications with unmanned aerial vehicles: opportunities and challenges,'' \emph{IEEE Commun. Mag.}, vol. 54, no. 5, pp. 36-42, May 2016.


\bibitem{b2}
X. Lin \emph{et al.}, ``The sky is not the limit: LTE for unmanned aerial vehicles,'' \emph{IEEE Commun. Mag.}, vol. 56, no. 4, pp. 204-210, Apr. 2018.

\bibitem{b1}
Y. Zeng, J. Lyu and R. Zhang, ``Cellular-connected UAV: potential, challenges, and promising technologies,'' \emph{IEEE Wireless Commun.}, vol. 26, no. 1, pp. 120-127, Feb. 2019.


\bibitem{b3}
W. Mei, Q. Wu and R. Zhang, ``Cellular-connected UAV: uplink association, power control and tnterference coordination,'' \emph{IEEE Trans. Wireless Commun.}, vol. 18, no. 11, pp. 5380-5393, Nov. 2019.

\bibitem{b4}
B. Van Der Bergh, A. Chiumento, and S. Pollin, ``LTE in the sky: trading off propagation benefits with interference costs for aerial nodes,'' \emph{ IEEE Commun. Mag.}, vol. 54, no. 5, pp. 44–50, May 2016.

\bibitem{b5}
A. Fotouhi \emph{et al.}, ``Survey on UAV cellular communications: practical aspects, standardization advancements, regulation, and security challenges,'' \emph{IEEE Commun. Surveys Tuts.}, vol. 21, no. 4, pp. 3417-3442, Fourthquarter 2019.

\bibitem{b6}
M. M. Azari, F. Rosas, A. Chiumento and S. Pollin, ``Coexistence of terrestrial and aerial users in cellular networks,'' in \emph{Proc. IEEE Globecom Workshops (GC Wkshps)}, Singapore, 2017, pp. 1-6.

\bibitem{b7}
V. V. Chetlur and H. S. Dhillon, ``Downlink coverage analysis for a finite 3-d wireless network of unmanned aerial vehicles,'' \emph{IEEE Trans. Wireless Commun.},  vol. 65, no. 10, pp. 4543-4558, Oct. 2017.

\bibitem{b8}
A. K. Gupta, H. S. Dhillon, S. Vishwanath and J. G. Andrews, ``Downlink coverage probability in MIMO hetnets with flexible cell selection,'' in \emph{Proc. IEEE Globecom.}, Austin, TX, Dec. 2014, pp. 1534-1539.

\bibitem{b17}
Z. Cui, C. Briso-Rodr\'iguez, K. Guan, \.I. G\"uven\c c and Z. Zhong, ``Wideband air-to-ground channel characterization for multiple propagation environments,'' \emph{IEEE Antennas. Wireless Propag. Lett.}, vol. 19, no. 9, pp. 1634-1638, Sept. 2020.

\bibitem{a2}
Z. Cui, C. Briso-Rodr\'iguez, K. Guan, Z. Zhong and F. Quitin, ``Multi-frequency air-to-ground channel measurements and analysis for UAV communication systems,'' \emph{IEEE Access}, vol. 8, pp. 110565-110574, Jun. 2020.

\bibitem{a3}
R. Amer, W. Saad and N. Marchetti, ``Mobility in the sky: performance and mobility analysis for cellular-connected UAVs,'' \emph{IEEE Trans. Commun.}, vol. 68, no. 5, pp. 3229-3246, May 2020.

\bibitem{a4}
X. Yu, J. Zhang, R. Schober and K. B. Letaief, ``A tractable framework for coverage analysis of cellular-connected UAV networks,'' in \emph{Proc. IEEE ICC Workshops}, Shanghai, China, Jun. 2019, pp. 1-6.

\bibitem{b9}
3GPP TR 36.777, ``Study on enhanced LTE support for aerial vehicles
(Release 15),'' V15.0.0, Dec. 2017

\bibitem{b10}
3GPP, ``Technical specification group radio access network; study on 3D channel model for LTE (Release 12),'' 3rd Generation Partnership Project (3GPP), TR 36.873 V12.2.0, June 2015.

\bibitem{a5}
3GPP, ``Study on channel model for frequencies from 0.5 to 100 GHz,'' 3rd Generation Partnership Project (3GPP), Tech. Rep. TR 38.901 V16.0.0 Release 16, Oct. 2019.

\bibitem{b13}
Z. Cui, K. Guan, C. Briso-Rodr\'iguez, B. Ai and Z. Zhong, ``Frequency-dependent line-of-sight probability modeling in built-up environments,'' \emph{IEEE Internet Things J.}, vol. 7, no. 1, pp. 699-709, Jan. 2020.

\bibitem{b11}
J. Yang, M. Ding, G. Mao, Z. Lin, D. Zhang and T. H. Luan, ``Optimal base station antenna downtilt in downlink cellular networks,'' \emph{IEEE Trans. Wireless Commun.}, vol. 18, no. 3, pp. 1779-1791, Mar. 2019.


\bibitem{b14}
M. Alzenad and H. Yanikomeroglu, ``Coverage and rate analysis for vertical heterogeneous networks (VHetNets),'' \emph{IEEE Trans. Wireless Commun.}, vol. 18, no. 12, pp. 5643-5657, Dec. 2019.

\bibitem{b18}
L. Zhou, Z. Yang, S. Zhou and W. Zhang, ``Coverage probability analysis of UAV cellular networks in urban environments,'' in \emph{Proc. IEEE ICC Workshops}, Kansas City, MO, USA, May 2018, pp. 1-6.


\bibitem{b15}
J. Barrett, ``5G spectrum bands,'' Global Mobile Suppliers Association, Feb. 2017. [Online]. Available: https://gsacom.com/5g-spectrum-bands/.

\end{thebibliography}
\end{document}